\documentclass[%
 reprint,
 groupedaddress,
 showpacs,preprintnumbers,
 nofootinbib,
 amsmath,amssymb,
 aps,
 pra,
]{revtex4-1}

\usepackage{dcolumn}% Align table columns on decimal point
\usepackage{bm}% bold math
\usepackage{amsthm}
\usepackage{dsfont}
\usepackage{verbatim}

\usepackage[pdfstartview={FitH top}]{hyperref} %,pdfpagemode=None
\hypersetup{
    colorlinks,
    citecolor=black,
    filecolor=black,
    linkcolor=black,
    urlcolor=blue
}

\newcommand {\be}{\begin{equation}}
\newcommand {\ee}{\end{equation}}

\theoremstyle{definition}

\newtheorem{remark}{\it Remark}
\newtheorem{example}{\it Example}

\newcommand{\Tr}{\mathop{\mathrm{Tr}}\nolimits}

\newcommand{\diag}{\mathop{\mathrm{diag}}\nolimits}
\allowdisplaybreaks
\begin{document}

\title{Entanglement in composite bosons realized by deformed oscillators}

\author{A.M. Gavrilik}
\affiliation{Bogolyubov Institute for Theoretical Physics of NAS of Ukraine, 14b Metrolohichna str., Kyiv 03680, Ukraine}
\email{omgavr@bitp.kiev.ua}
%\address{}

\author{Yu.A. Mishchenko}
\affiliation{Bogolyubov Institute for Theoretical Physics of NAS of Ukraine, 14b Metrolohichna str., Kyiv 03680, Ukraine}

\begin{abstract}
Composite bosons (or quasibosons), as recently proven, are realizable by deformed oscillators and due to that can be effectively treated as particles of nonstandard statistics (deformed bosons). This enables us to study quasiboson states and their inter-component entanglement aspects using the well developed formalism of deformed oscillators. We prove that the internal entanglement characteristics for single two-component quasiboson are determined completely by the parameter(s) of deformation. The bipartite entanglement characteristics are generalized and calculated for arbitrary multi-quasiboson (Fock, coherent etc.) states and expressed through deformation parameter.
\end{abstract}
%\keywords{}

\maketitle

%\tableofcontents

\section{Introduction}

Composite bosons are encountered in diverse branches of quantum physics ranging from (sub)nuclear (mesons, higgson, light even nuclei) and condensed matter (excitons, cooperons) physics to atomic, molecular and quantum information (biphotons) theory. As composite bosons components are usually entangled, generic bipartite entangled states~\cite{Horodecki,Tichy_Rev} used in quantum information applications~\cite{Nielsen_Chuang} can be viewed also as quasibosonic states, local or nonlocal. The treatment of many quasibosons system is quite involved due to their substructure. It can be performed directly using polylinear representation of quasibosons' creation/annihilation operators by the constituents' ones, like in~\cite{Combescot_ManyBody}. In some limit quasibosons approach pure bosons; as shown in~\cite{Law,Chudzicki}, a good criterium to estimate the extent of approaching can be presented in terms of the entanglement of quasiboson constituents.

Also, quasibosons can be modeled~\cite{Avan,GKM,GKM2} by deformed bosons (deformed oscillators~\cite{Bonatsos}), a nonlinear generalization of usual bosons. That enables to cover a wide class of quasiboson wavefunctions, and to encapsulate the peculiarities of internal structure into deformation parameters. Note that the theory of deformed oscillators was actively developed in the last two decades, including both $q$-deformations and generalized deformed oscillator~\cite{Man'ko,Meljanac}. The deformed oscillators were efficiently used to treat nucleons and quarks~\cite{Sviratcheva}, molecules~\cite{Bonatsos_Mol}, excitons~\cite{Bagheri_Harouni}, phonon systems~\cite{R-Monteiro}, Bose-Einstein condensation~\cite{Algin_BECond}, particles of non-standard statistics~\cite{Greenberg_Statist}, interacting particles systems~\cite{Scarfone_Interact}. Besides, many-particle systems of (mainly~$q$-) deformed bosons were explored in the context  of thermodynamics~\cite{Algin_Therm,Lavagno_Therm} and correlations~\cite{Anchishkin,Adamska,GR_EPJA}.

In this paper we explore the role of entanglement for a special class of {\it bipartite} quasibosons, consisting of two fermions or two bosons, which admit realization, at the level of creation/annihilation operators, by deformed bosons. The approach developed in~\cite{GKM,GKM2} allows us here to establish direct connection between the (degree/measure of the) entanglement within a composite boson and the (parameter of the) deformation. We calculate the measures of bipartite entanglement also for multi-quasibosonic states, e.g. Fock states or coherent state, and express them through the deformation parameter.

\section{Quasiboson as entangled bipartite system}

Bipartite quasiboson can be viewed as entangled system regarding to its components.  Quasiboson state vector in $\alpha$-th mode, as an element of tensor product of components' Hilbert spaces, is expressed in the general form
\begin{equation}\label{1state}
|\Psi_{\alpha}\rangle \!=\! \sum_{\mu\nu}\Phi_{\alpha}^{\mu\nu} |a_{\mu}\rangle\otimes|b_{\nu}\rangle,\ \ |a_{\mu}\rangle\!\equiv a^{\dag}_{\mu}|0\rangle,\ |b_{\nu}\rangle\!\equiv b^{\dag}_{\nu}|0\rangle
\end{equation}
with the creation operators $a^{\dag}_{\mu}$, $b^{\dag}_{\nu}$ of the constituents (both pure fermions, or both pure bosons) and complex $d_a\times d_b$ matrix $\parallel\!\!\Phi_{\alpha}^{\mu\nu}\!\!\parallel$. For such bipartite state, the corresponding Schmidt decomposition~\cite{Nielsen_Chuang} does hold:
\begin{equation}\label{decomposition}
|\Psi_{\alpha}\rangle = \sum_{k=1}^{\min(d_a,d_b)} \lambda^{\alpha}_k |v^{\alpha}_k\rangle\otimes|w^{\alpha}_k\rangle
\end{equation}
where $\sum_k(\lambda^{\alpha}_k)^2=1$ that implies normalized states $|\Psi_{\alpha}\rangle$. The set of real coefficients $\lambda^{\alpha}_k$ is unique up to permutations for any fixed state $|\Psi_{\alpha}\rangle$. Each ingredient of Schmidt decomposition (i.e. $\lambda^{\alpha}_k$, $|v^{\alpha}_k\rangle$, $|w^{\alpha}_k\rangle$) depends on the state $|\Psi_{\alpha}\rangle$. If the number of indices $k$ for which $\lambda^{\alpha}_k\neq 0$ is greater than one then the state  $|\Psi_{\alpha}\rangle$ is entangled, otherwise it is separable.

\section{Quasibosons realized by deformed oscillators}

In this work we deal with the special class of quasibosons, i.e. those which can be represented by deformed oscillators. Detailed description of such an approach is given in \cite{GKM,GKM2}. Here we only sketch some results on the realization of quasibosons by deformed oscillators, necessary for subsequent analysis. The creation operator for the  composite "fermion+fermion" or "boson+boson" type quasiboson in the state (\ref{1state}) is taken, see~\cite{Avan,GKM,GKM2}, as
\begin{equation}\label{anzats}
A^{\dag}_{\alpha} = \sum_{\mu\nu}\Phi_{\alpha}^{\mu\nu} a^{\dag}_{\mu}b^{\dag}_{\nu},\quad\text{so that}\quad |\Psi_{\alpha}\rangle=A^{\dag}_{\alpha}|0\rangle.
\end{equation}
We intend to find the algebra, generated by $A_{\alpha}$, $A^{\dag}_{\alpha}$ and $N_{\alpha}\equiv\phi^{-1}(A^{\dag}_{\alpha}A_{\alpha})$,
corresponding to multimode system of commuting deformed oscillators, each mode being given by the structure function $\phi(N)$~\cite{Meljanac}. Then, on the subspace of {\it all the quasibosons states}, i.e. linear span of $\{(A^{\dag}_{\gamma_1})^{m_1}\cdot...\cdot (A^{\dag}_{\gamma_n})^{m_n}|0\rangle|n\in\mathds{N}\}$, there should hold
\begin{subequations}
\begin{align}
&[A_{\alpha},A^{\dag}_{\beta}]\equiv -\epsilon \Delta_{\alpha\beta} = 0 \quad\text{for}\quad \alpha\neq\beta,\label{eq1}\\
&[N_{\alpha},A^{\dag}_{\alpha}]= A^{\dag}_{\alpha},\quad [N_{\alpha},A_{\alpha}]= -A_{\alpha},\label{eq2}\\
&[A_{\alpha},A^{\dag}_{\alpha}]\equiv 1-\epsilon\Delta_{\alpha\alpha} = \phi(N_{\alpha}+1)-\phi(N_{\alpha}),\label{eq3}
\end{align}
\end{subequations}
%============================Page 2=====================================
where $\epsilon=+1$ or $\epsilon=-1$ if the constituents are respectively fermions or bosons\footnote[1]{Although the papers \cite{GKM,GKM2} deal with the case of fermionic constituents, the treatment of quasibosons built from pure bosonic constituents proceeds along similar lines, and the resulting formulas in the both cases can be given in a unified way using $\epsilon$.}. Note that in (\ref{eq1}), (\ref{eq3}) we have used the orthonormality condition $\Tr(\Phi_{\alpha}\Phi_{\beta}^{\dag})=\delta_{\alpha\beta}$ and denoted by $\Delta_{\alpha\beta}$ the deviation from standard Bose commutation relations, namely
\begin{equation*}
\Delta_{\alpha\beta} = \sum_{\mu\mu'}(\Phi_{\beta}{\Phi}^{\dag}_{\alpha})^{\mu'\mu}
a^{\dag}_{\mu'}a_{\mu} + \sum_{\nu\nu'}({\Phi}^{\dag}_{\alpha}
\Phi_{\beta})^{\nu\nu'} b^{\dag}_{\nu'}b_{\nu}.
\end{equation*}
Equality (\ref{eq1}) guarantees independence of different modes of quasibosons. It is equivalent, as proven in~\cite{GKM,GKM2}, to the following relation involving the matrices $\Phi$:
\begin{equation}
\Phi_{\beta}\Phi^{\dag}_{\alpha}\Phi_{\gamma}+
\Phi_{\gamma}\Phi^{\dag}_{\alpha}\Phi_{\beta}=0,\quad
\alpha\neq\beta.\label{req1}
\end{equation}
Basing on (\ref{eq2}) and (\ref{eq3}), rewritten as $F\!\equiv\! \epsilon\Delta_{\alpha\alpha}\!-1 \!+
\phi(N_{\alpha}\!+\!1)\!-\!\phi(N_{\alpha})=0$, we calculate the $n$-th commutator
\begin{multline}\label{n-commut}
\bigl[...\bigl[F,A^{\dag}_{\alpha}\bigr]... A^{\dag}_{\alpha}\bigr]
= 2\delta_{n,1} \sum_{\mu\nu} (\Phi_{\alpha}\Phi^{\dag}_{\alpha}\Phi_{\alpha})^{\mu\nu} a^{\dag}_{\mu}b^{\dag}_{\nu} +\\
+ (A^{\dag}_{\alpha})^n \biggl\{\sum_{k=0}^{n+1}
C^k_{n+1}\phi(N_{\alpha}\!+\!k) (-1)^{n+1-k}\biggr\}
\mathop{=}\limits^{(\ref{eq3})}0.
\end{multline}
Eq. (\ref{n-commut}) yields one more requirement for matrices~$\Phi_{\alpha}$,
\begin{equation}\label{req2}
\Phi_{\alpha}\Phi^{\dag}_{\alpha}\Phi_{\alpha} = \frac{f}{2} \Phi_{\alpha}
\end{equation}
with the (deformation) parameter $f \equiv \epsilon\!\cdot\!(2-\phi(2))=2\Tr(\Phi^{\dag}_{\alpha}\Phi_{\alpha}\Phi^{\dag}_{\alpha}\Phi_{\alpha})$ shared by all the modes $\alpha$, and the recurrence relation for the structure function $\phi$:
\begin{equation}
\phi(n+1) = \sum_{k=0}^{n} (-1)^{n-k} C^k_{n+1}\phi(k), \quad n\geq
2.\label{recurr1}
\end{equation}
Its general solution, obeying the initial conditions $\phi(1)=1$ and $\phi(2)\!=\!2\!-\!\epsilon f$, can be written in the form
\begin{equation}
\phi(n)=\Bigl(1+\epsilon \frac{f}{2}\Bigr)n - \epsilon \frac{f}{2}n^2.\label{solution1}
\end{equation}
General solution for matrices $\Phi_{\alpha}$ which satisfy equations (\ref{req1}), (\ref{req2}) and the orthonormality condition was found in~\cite{GKM2}. It is expressed in the block-diagonal form, using arbitrary fixed unitary matrices $U_1(d_a)$ and $U_2(d_b)$:
\begin{equation}
\Phi_{\alpha}\!=\!U_1(d_a)
\diag\Bigl\{0..0,\sqrt{f/2}\, U_{\alpha}(m),0..0\Bigr\}U^{\dag}_2(d_b).\label{gen_solution}
\end{equation}
Here the $m\times m$ block $\sqrt{f/2}\, U_{\alpha}(m)$ is at its $\alpha$-th place, and does not intersect with the corresponding blocks of other matrices $\Phi_{\beta}$. Also, during the derivation we have found that the deformation parameter $f$ takes only discrete values: $f/2=1/m$.

\section{Measures of the entanglement within quasiboson}

Our consideration, as indicated, deals with quasibosons which are realized by deformed oscillators. So, according to (\ref{gen_solution}) we have the following explicit form of~$\Phi_{\alpha}$:
\begin{align*}
&\Phi_{\alpha}
=U_1(d_a) \diag\Bigl\{0..0,\sqrt{f/2}\, U_{\alpha}(m),0..0\Bigr\}U^{\dag}_2(d_b) =\\
&=\! U_1(d_a) \diag\Bigl\{\!0,\!\sqrt{\!f\!/2}\,\mathds{1}_m,0\!\Bigr\}
\diag\Bigl\{\!\mathds{1},U_{\alpha}(m),\mathds{1}\!\Bigr\}U^{\dag}_2(d_b)\!=\\
&\qquad\qquad
=U_1(d_a) \diag\Bigl\{0..0,\sqrt{f/2}\ \mathds{1}_m,0..0\Bigr\}\tilde{U}_2(d_b)
\end{align*}
where $ \tilde{U}_2(d_b) \equiv \diag\bigl\{\mathds{1},U_{\alpha}(m),\mathds{1}\bigr\}U^{\dag}_2(d_b)$. Using  the latter we rewrite the state $|\Psi_{\alpha}\rangle$ in the form of Schmidt decomposition (\ref{decomposition}), namely
\begin{align}
&|\Psi_{\alpha}\rangle \!=\! \sum_{k=1}^m\! \frac{1}{\sqrt{m}}
|v^{\alpha}_k\rangle\!\otimes\!|w^{\alpha}_k\rangle,\
|v^{\alpha}_k\rangle\!\!=\!U_1^{\mu k}|a_{\mu}\rangle,
|w^{\alpha}_k\rangle\!\!=\!\tilde{U}_2^{k\nu}|b_{\nu}\rangle,\nonumber\\
&\lambda^{\alpha}_k=\lambda=\sqrt{f/2}=1/\sqrt{m}.\label{lambda_f}
\end{align}
As seen, it is the parameter $m$ (linked with the parameter $f$ of deformation as $\frac{f}{2}=\frac1m$) that determines the Schmidt coefficients $\lambda^{\alpha}_k$. Then, the degree of entanglement is characterized by the following quantities (see e.g. reviews \cite{Horodecki,Tichy_Rev} for their definition):
\begin{itemize}
\item Schmidt rank, the number of non-vanishing Schmidt coefficients $\lambda^{\alpha}_k$, here equal to $m$;
\item Schmidt number (and the purity $P$ of subsystem)
\begin{equation} \label{char2}
K=\biggl[\sum_k (\lambda^{\alpha}_k)^4\biggr]^{-1} = 1/P = m;
\end{equation}
\item Entanglement entropy
\begin{equation}\label{char3}
S_{\rm entang}=-\sum_k (\lambda^{\alpha}_k)^2 \ln
(\lambda^{\alpha}_k)^2 = \ln(m);
\end{equation}
\item Concurrence
\begin{equation}\label{char4}
C=\biggl[\frac{m}{m-1}\Bigl(1-\sum_k (\lambda^{\alpha}_k)^4\Bigr)\biggr]^{1/2}= 1.
\end{equation}
\end{itemize}
\begin{remark}
The inequalities obtained in \cite{Chudzicki} in our terms take the form (see \cite{Law} for the definition of $\chi_N$)
\begin{equation}
1-N/m \le \chi_{N+1}/\chi_N \le 1-1/m.
\end{equation}
\end{remark}
>From (\ref{char2})-(\ref{char4}) we see that the greater is $m$, the greater is entanglement of fermions (bosons) in the quasiboson, and the smaller is the deformation parameter $f$. Hence, in view of (\ref{solution1}), a strongly entangled quasiboson approaches pure boson, for small enough quantum numbers $N$:
\begin{equation*}
\phi(N)\approx \phi_{boson}(N)\equiv N,  \quad N\ll m, \  \ m\gg 1.
\end{equation*}
Note however that quasibosons cannot, especially for large $N$, be strictly realized as pure bosons \cite{GKM2}.

Using Schmidt decomposition we can write the following expression for quasiboson creation operator:
\begin{equation}\label{A_dec}
A_{\alpha}^{\dag} = \sum_{k=1}^m \lambda^{\alpha}_k
v^{\alpha\dag}_k w^{\alpha\dag}_k,\ \ v^{\alpha\dag}_k\equiv 
U_1^{\mu k}a^{\dag}_{\mu},\,
w^{\alpha\dag}_k \equiv \tilde{U}_2^{k\nu}b^{\dag}_{\nu}.
\end{equation}
It is worth noting that, due to unitarity of $U_1$ and $\tilde{U}_2$, the operators $v_k$ and $w_k$ obey the same commutation relations as $a_{\mu}$ and $b_{\nu}$, i.e. (the same hold for $w_k$):
\begin{align*}
&\{v^{\alpha}_k,v^{\beta\dag}_{k'}\}=\delta_{kk'}\delta_{\alpha\beta},& &\{v^{\alpha\dag}_k,v^{\beta\dag}_{k'}\}=0, & &\text{if}\ \ \epsilon=+1,\\
&[v^{\alpha}_k,v^{\beta\dag}_{k'}]=\delta_{kk'}\delta_{\alpha\beta},& &[v^{\alpha\dag}_k,v^{\beta\dag}_{k'}]=0, & &\text{if}\ \ \epsilon=-1.
\end{align*}
Then we can rewrite the deviation $\Delta_{\alpha\alpha}$ as
\begin{equation*}
\Delta_{\alpha\alpha}=\sum_{k=1}^m
(\lambda^{\alpha}_k)^2(v^{\alpha\dag}_kv^{\alpha}_k+w^{\alpha\dag}_kw^{\alpha}_k)
\end{equation*}
in terms of $v^{\alpha}_k$ and $w^{\alpha}_k$ instead of $a_{\mu}$ and $b_{\nu}$.
%============================Page 3=====================================

Now let us convert some of the statements into the ``density matrix'' formalism. Using (\ref{decomposition}), the density matrix for our composite system reads
\begin{equation*}
\rho_{\alpha} = \sum_{k,j} \lambda^{\alpha}_k\lambda^{\alpha}_j
|v^{\alpha}_k\rangle\langle v^{\alpha}_j|\otimes
|w^{\alpha}_k\rangle\langle w^{\alpha}_j|.
\end{equation*}
With (\ref{lambda_f}) this yields the density matrices for subsystems:
\begin{align*}
&\rho_{\alpha}^{(a)} = \Tr_{b} \rho_{\alpha} =
\sum_{k=1}^m \frac1m |v^{\alpha}_k\rangle\langle v^{\alpha}_k|,\\
&\rho_{\alpha}^{(b)} = \Tr_{a} \rho_{\alpha} = \sum_{k=1}^m
\frac1m |w^{\alpha}_k\rangle\langle w^{\alpha}_k|.
\end{align*}
Thus we have obtained mixed states for the subsystems (for $m>1$). That means that the whole composite system is in entangled state. So, in terms of the density matrix we equivalently (compare with (\ref{char2}) and (\ref{char3})) find:
\begin{itemize}
\item the Schmidt number
\begin{equation*}
K=1/\Tr (\rho_{\alpha}^{(a)})^2 = 1/\Tr (\rho_{\alpha}^{(b)})^2 = m;
\end{equation*}
\item the entanglement entropy
\begin{equation*}
S_{\rm entang} \!=\! -\!\Tr_{a} \rho_{\alpha}^{(a)}\ln \rho_{\alpha}^{(a)} \!=\! -\!\Tr_{b} \rho_{\alpha}^{(b)}\ln \rho_{\alpha}^{(b)} \!=\! \ln(m).
\end{equation*}
\end{itemize}
Recall that $m$ originates from (and determines) the considered deformation, see (\ref{solution1})-(\ref{lambda_f}), used to realize quasibosons (\ref{anzats}). At the same time it appears to be the measure of entanglement for the state of one quasiboson.

\section{Entanglement in multi-quasiboson states}

Unlike previous case where the subsystems were elementary, it is possible to introduce (in a natural way) the bipartite entanglement for {\it multi-quasiboson} states using instead of elementary subsystems their tensor products. So, let us now extend the previous treatment to the case of most general multi-quasibosons states
\begin{equation}\label{init_state}
|\Psi\rangle = \sum\limits_{\{m_{\gamma}\}} \Psi\bigl(\{m_{\gamma}\}\bigr) (A^{\dag}_{\gamma_1})^{m_{\gamma_1}}\cdot...\cdot(A^{\dag}_{\gamma_D})^{m_{\gamma_D}}|0\rangle
\end{equation}
where $\{m_\gamma\}$ is a set of occupation numbers for all quasibosonic modes $\gamma\!\in\!\!\{\gamma_1$,...,$\gamma_D\}$; $\Psi\bigl(\!\{m_{\gamma}\!\}\!\bigr)\!\equiv\!\Psi(m_{\gamma_1},...,m_{\gamma_D})$ is a state wavefunction in second quantized representation. The summation in (\ref{init_state}) runs over all possible sets of occupation numbers $\{m_{\gamma}\}$ for quasibosons. The states (\ref{init_state}) are
considered as bipartite entangled with respect to $a$- and $b$-subsystems. That is, the first or $a$-subsystem collects all the $a$-type fermions (bosons), and likewise $b$-subsystem collects all the $b$-type fermions (bosons).
\begin{remark}\label{rem2}
Since for $\epsilon=+1$ the operators $A^{\dag}_{\gamma}$, as one can show, are nilpotent of order $m+1$, we have the restriction $m_{\gamma}\le m$ for this case. Then,  the sum in (\ref{init_state}) is finite if the set of modes $\alpha$ is finite. However, for the case $\epsilon=-1$ the operators $A^{\dag}_{\gamma}$ are not nilpotent, and the sum in (\ref{init_state}) can in general be infinite, like for coherent states.
\end{remark}
Let us write using (\ref{A_dec}) the building block of this state in the form
\begin{equation}\label{m-state}
(A^{\dag}_{\gamma})^{m_{\gamma}} \!=\! \Bigl(\frac1m\Bigr)^{m_{\gamma}/2} \hspace{-1.5mm}m_{\gamma}! \hspace{-4mm}\sum\limits_{k_1\le...\le k_{m_{\gamma}}}\hspace{-5mm} \frac{v^{\gamma\dag}_{k_1}...v^{\gamma\dag}_{k_{m_{\gamma}}}\!\!}{R_{k_1...k_{m_\gamma}}}\!\otimes\!\! \frac{w^{\gamma\dag}_{k_{m_{\gamma}}}...w^{\gamma\dag}_{k_1}\!\!}{R_{k_1...k_{m_\gamma}}}
\end{equation}
which corresponds to Schmidt decomposition; $R_{k_1...k_{m_\gamma}}\equiv\bigl|v^{\gamma\dag}_{k_1}...v^{\gamma\dag}_{k_{m_{\gamma}}}\!|0\rangle\bigr| = \bigl|w^{\gamma\dag}_{k_{m_{\gamma}}}...w^{\gamma\dag}_{k_1}|0\rangle\bigr|$ is normalizing coefficient. Introducing multiindex notations
\begin{equation*}
k^{\gamma}\!=\!(k_1,...,k_{m_{\gamma}}),\ \ V^{\gamma\dag}_{k^{\gamma}}\equiv \frac{v^{\gamma\dag}_{k_1}...v^{\gamma\dag}_{k_{m_{\gamma}}}\!\!}{R_{k_1...k_{m_\gamma}}},\ \
W^{\gamma\dag}_{k^{\gamma}}\equiv \frac{w^{\gamma\dag}_{k_{m_{\gamma}}}...w^{\gamma\dag}_{k_1}\!\!}{R_{k_1...k_{m_\gamma}}},
\end{equation*}
we rewrite the operator (\ref{m-state}) more compactly:
\begin{equation*}
(A^{\dag}_{\gamma})^{m_{\gamma}} = \Bigl(\frac1m\Bigr)^{m_{\gamma}/2} m_{\gamma}! \sum\limits_{\text{ordered }k^{\gamma}} V^{\gamma\dag}_{k^{\gamma}}\otimes W^{\gamma\dag}_{k^{\gamma}}.
\end{equation*}
Then the initial state (\ref{init_state}) is written as extended Schmidt decomposition, cf.~(\ref{decomposition}), (with summation over the set of ordered $k^{\gamma_1},...,k^{\gamma_D}$):
\begin{multline*}
|\Psi\rangle \!=\! \!\sum\limits_{\{m_{\gamma}\}} \sum\limits_{k^{\gamma_1}\!\!,...,k^{\gamma_D}}\hspace{-4mm}\Lambda_{k^{\gamma_1}\!,...,k^{\gamma_D}} \cdot V^{\gamma_1\dag}_{k^{\gamma_1}}...V^{\gamma_D\dag}_{k^{\gamma_D}}\otimes W^{\gamma_D\dag}_{k^{\gamma_D}}...W^{\gamma_1\dag}_{k^{\gamma_1}}|0\rangle,\\
\Lambda_{k^{\gamma_1},...,k^{\gamma_D}} = \Psi\bigl(\{m_{\gamma}\}\bigr) \Bigl(\frac1m\Bigr)^{(m_{\gamma_1}+...+m_{\gamma_D})/2} m_{\gamma_1}!... m_{\gamma_D}!,
\end{multline*}
with the coefficients $\Lambda_{k^{\gamma_1},...,k^{\gamma_D}}$. On its base we calculate the characteristics of bipartite entanglement:
\begin{itemize}
\item Schmidt number (defined as in (\ref{char2}))
\begin{equation}
K \!\!=\!\! \Biggl[\sum\limits_{\{m_{\gamma}\!\}}\! \bigl|\Psi\bigl(\!\{m_{\gamma}\!\}\!\bigr)\bigr|^4 \Bigl(\frac1m\Bigr)^{2\!\sum\limits_{j=1}^D \!m_{\gamma_j}} \!\!\! \prod\limits_{j=1}^D (m_{\gamma_j}!)^4 \!N_m^{m_{\gamma_j}}\!\Biggr]^{\!-\!1}\!;\!\label{mchar2}
\end{equation}
\item Entanglement entropy (defined as in (\ref{char3}))
\begin{multline}
S_{\rm entang} \!=\! -\!\!\sum\limits_{\{m_{\gamma}\}} \!\!|\Psi\bigl(\!\{m_{\gamma}\!\}\!\bigr)|^2 \Bigl(\frac1m\Bigr)^{\sum\limits_{j=1}^D m_{\gamma_j}} \!\! \prod\limits_{j=1}^D (m_{\gamma_j}!)^2 N_m^{m_{\gamma_j}}\cdot\\
\cdot\ln \biggl[|\Psi\bigl(\{m_{\gamma}\}\bigr)|^2 \Bigl(\frac1m\Bigr)^{\sum\limits_{j=1}^D m_{\gamma_j}} \prod\limits_{j=1}^D (m_{\gamma_j}!)^2\biggr].\label{mchar3}
\end{multline}
\end{itemize}
Here $N_m^{m_{\gamma}} = C_m^{m_{\gamma}}$ for quasibosons composed of two pure fermions, and $N_m^{m_{\gamma}} = C_{m+m_{\gamma}-1}^{m-1}$ for quasibosons composed of two pure bosons ($C_m^n$ is binomial coefficient).
%==================Example 1===========================
\begin{example}\label{ex1}
Let us calculate the characteristics of entanglement for the normalized multi-quasibosonic Fock state $[\phi(m_{\alpha})!]^{-1/2}(A^{\dag}_{\alpha})^{m_{\alpha}}|0\rangle$ for a fixed mode $\alpha$ (here the $\phi$-factorial means $\phi(n)!\mathop{=}\limits^{def} \phi(1)\cdot...\cdot\phi(n)$). For such a state the amplitude $\Psi$ for the configuration  $\{m_{\gamma}=m_{\alpha},\gamma=\alpha;\ m_{\gamma}=0,\gamma\neq\alpha\}$ of occupation numbers is $\Psi =[\phi(m_{\alpha})!]^{-1/2}$, while for other configurations $\Psi=0$. Taking into account that, for our structure function (\ref{solution1}), $\phi(n)!|_{\epsilon=+1}=(n!)^2C_m^n/m^n$ and $\phi(n)!|_{\epsilon=-1}=(n!)^2C_{m+n-1}^n/m^n$ we obtain:
\begin{equation}\label{Fock1}
\begin{aligned}
K_{\epsilon=+1}\!=\! C_m^{m_{\alpha}} &, \ K_{\epsilon=-1}\!=\! C_{m+m_{\alpha}-1}^{m_{\alpha}};\\
S_{\rm entang}|_{\epsilon=+\!1}\!=\! \ln C_m^{m_{\alpha}} &, \ S_{\rm entang}|_{\epsilon=-\!1}\!=\! \ln C_{m\!+\!m_{\alpha}\!-1}^{m_{\alpha}}.
\end{aligned}
\end{equation}
\end{example}
%==================Example 2===========================
\begin{example}\label{ex2}
Likewise we can consider $n$-quasiboson state when all quasibosons are in different modes:
\begin{equation*}
|\Psi\rangle = A^{\dag}_{\gamma_1}\cdot...\cdot A^{\dag}_{\gamma_n}|0\rangle,\quad\gamma_i\neq \gamma_j,\ i\neq j,\ i,j=1,...,n.
\end{equation*}
Performing similar calculations we find:
\begin{equation}\label{Fock2}
\begin{aligned}
&K_{\epsilon=+1} =  K_{\epsilon=-1}= m^n;\\
&S_{\rm entang}|_{\epsilon=+1} =  S_{\rm entang}|_{\epsilon=-1}= n\ln (m).
\end{aligned}
\end{equation}
\end{example}
%====================Page 4====================================
\begin{remark}
Notice that in (\ref{char2}), (\ref{char3}) and in the Examples \ref{ex1} and \ref{ex2} we have simple rule: the entanglement entropy is $S_{\rm entang}=\ln K$.
\end{remark}
%==================Example 3===========================
\begin{example}\label{ex3}
It is of interest to calculate the entanglement characteristics for a {\it quasiboson's coherent state}, which exists only if quasiboson consists of two bosons (when $\epsilon=-1$, see Remark \ref{rem2}), like biphoton.  So, the coherent state for single quasibosonic mode $\alpha$ (recall that $A_{\alpha}|\Psi_{\alpha}\rangle=\mathcal{A}_{\alpha}|\Psi_{\alpha}\rangle$, and $A_{\alpha}$ is defined by (\ref{anzats})) can be expressed as the series (we drop subscript $\alpha$ in $\mathcal{A}_{\alpha}$)
\begin{align*}
&|\Psi_{\alpha}\rangle \!=\! \tilde{C}(\mathcal{A};m) \sum\limits_{n=0}^\infty \frac{\mathcal{A}^n}{\phi(n)!} (A^{\dag}_{\alpha})^n|0\rangle,\\
&\tilde{C}(\mathcal{A};m) \!=\! \biggl(\sum\limits_{n=0}^\infty \frac{|\mathcal{A}|^{2n}}{\phi(n)!}\biggr)^{-1/2} \!\!=\! \biggl[\frac{(m\!-\!1)! I_{m-1}(z)}{(z/2)^{m-1}}\biggr]^{-\frac12}= \\
&\qquad\quad\ \, = \mathrm{e}^{-|\mathcal{A}|^2/2}\Bigl[1+\frac14 \frac{|\mathcal{A}|^4}{m}+...\Bigr],\ \ \ z=2\sqrt{m}|\mathcal{A}|,
\end{align*}
where $I_{m-1}(z)$ is the modified Bessel function of order $m-1$. For this state we obtain
\begin{align}
&K=\tilde{C}^{-4} \left[\sum\limits_{n=0}^\infty (C^n_{n+m-1})^{-3} \frac{(|\mathcal{A}|^2m)^{2n}}{(n!)^4}\right]^{-1} = \nonumber\\
&=\!\tilde{C}^{-\!4}\! / {}_0\!F\!_3(m,\!m,\!m;\!|\mathcal{A}|^4m^2) \!=\! \mathrm{e}^{2|\mathcal{A}|^2}\!\Bigl[1\!-\!2\frac{|\mathcal{A}|^4\!\!}{m}\!+\!...\Bigr],\label{K_coherent}
\end{align}
where ${}_0\!F\!_3$ is the hypergeometric function, and also
\begin{align}
S_{\rm entang} &=\tilde{C}^2 \sum\limits_{n=0}^\infty \frac{(|\mathcal{A}|^2m)^n}{(n!)^2 C^n_{n+m-1}} \ln \biggl[\frac{(n!)^2(C_{n+m-1}^n)^2}{\tilde{C}^2 (|\mathcal{A}|^2m)^n}\biggr]=\nonumber\\
&= |\mathcal{A}|^2\Bigl[1-\frac12(1+|\mathcal{A}|^2)\frac{|\mathcal{A}|^2}{m}+...\Bigr]\ln \frac{m}{|\mathcal{A}|^2} + \nonumber\\ &+|\mathcal{A}|^2\Bigl[1+\Bigl(1-\frac{|\mathcal{A}|^2}{2}\Bigr)\frac{|\mathcal{A}|^2}{m}+...\Bigr].\label{S_coherent}
\end{align}
\end{example}
As we have shown, the concept of bipartite entanglement admits natural generalization to arbitrary {\it multi-quasibosonic} (including coherent) states. The degree of such entanglement in general depends on the wavefunction $\Psi\bigl(\{m_{\gamma}\}\bigr)$, see (\ref{init_state}), but for the ordinary Fock states we find: it is greater the greater is $m$ and (if $\epsilon=-1$) the number of quasibosons in the state.

\section{Summary}

We have proven that for composite bosons realized by deformed bosons (deformed oscillators) the entanglement characteristics are unambiguously, see~(\ref{char2})-(\ref{char4}), determined by the deformation parameter $f=2/m$, and conversely, the degree of the entanglement between constituents determines the strength of deformation. Thus, it is the internal entanglement of the constituents that reveals the physical meaning of the deformation parameter. Similar results, as seen from~(\ref{mchar2})-(\ref{mchar3}), have been established also for the multi-quasiboson states considered as a bipartite system, particularly for Fock states~(\ref{Fock1})-(\ref{Fock2}) and for the coherent states~(\ref{K_coherent})-(\ref{S_coherent}). It would be important to uncover physical implications of the established relation between entanglement and deformation, for those systems of particles of composite nature where deformed oscillators find efficient applications (e.g. pions~\cite{Gavrilik_Sigma,Ribeiro-Silva}, excitons~\cite{Bagheri_Harouni} and alike).

\begin{acknowledgments}
This research was  partially supported by the Special Program of the Division of Physics and Astronomy of NAS of Ukraine.
\end{acknowledgments}

%\nocite{*}

\bibliography{entanglement}

\end{document}